\documentclass{jfm}

\usepackage{graphicx}
\usepackage{newtxtext}
\usepackage{newtxmath}
\usepackage{natbib}
\usepackage{hyperref}
\hypersetup{
    colorlinks = true,
    urlcolor   = blue,
    citecolor  = black,
}

\newcommand{\RomanNumeralCaps}[1]
\linenumbers

\usepackage{amssymb}
\usepackage[dvipsnames]{xcolor}

\newcommand{\rme}{\mathrm{e}}
\newcommand{\rmi}{\mathrm{i}}
\newcommand{\bk}{\boldsymbol{k}}

\usepackage{soul}
\usepackage[normalem]{ulem}
\newcommand\redsout{\bgroup\markoverwith{\textcolor{red}{\rule[0.2ex]{2pt}{5pt}}}\ULon}

\usepackage{subcaption}
\usepackage{diagbox}
\usepackage{csquotes}

\title{Multiple states of two-dimensional turbulence above topography}

\author{Jiyang He\aff{1,2}
 \and Yan Wang\aff{1,2}
 \corresp{\email{yanwang@ust.hk}}
 }

\affiliation{\aff{1}Department of Ocean Science, The Hong Kong University of Science and Technology, Hong Kong, China\aff{2}Center for Ocean Research in Hong Kong and Macau, The Hong Kong University of Science and Technology, Hong Kong, China.}

\begin{document}
\maketitle

\begin{abstract}
The recent work of Siegelman \& Young (PNAS, vol. 120(44), 2023, pp. e2308018120) revealed two extreme states reached by the evolution of unforced and weakly-damped two-dimensional turbulence above random rough topography, separated by a critical kinetic energy $E_\#$. 
The low- and high-energy solutions correspond to topographically-locked and roaming vortices, surrounded by non-uniform and homogeneous background potential vorticity (PV), respectively.
However, we found that these phenomena are restricted to the particular intermediate length scale where the energy was initially injected into the system.
Through simulations initialised by injecting the energy at larger and smaller length scales, we found that the long-term state of topographic turbulence is also dependent on the initial length scale and thus the initial enstrophy.
If the initial length scale is comparable with the domain size, the long-term flow field resembles the minimum-enstrophy state proposed by Bretherton \& Haidvogel (J. Fluid Mech., vol.78(1), 1976, pp. 129-154), with very few topographically-locked vortices; the long-term enstrophy is quite close to the minimum value, especially when the energy is no larger than $E_\#$.
As the initial length scale becomes smaller, more vortices nucleate and become more mobile;
the long-term enstrophy increasingly deviates from the minimum value.
Simultaneously, the background PV tends to homogenization, even if the energy is below $E_\#$.
These results complement the phenomenology of topographic turbulence documented by Siegelman \& Young, by showing that the minimum-enstrophy and background PV homogenization states can be adequately approached by large- and small-scale initial fields, respectively, with relatively arbitrary energy.

\end{abstract}


\section{Introduction}
Two-dimensional, rapidly-rotating turbulence above topography (topographic turbulence hereafter) serves as an effective reduced model for studying large-scale oceanic motions over rough and sloping seafloor. 
This model has provided insight into the emergence of prevailing undercurrents over continental shelves and slopes \citep{WangStewart2018Slope} and persistent along-bathymetry flows above topographic depressions and seamounts \citep{solodochFormationAnticyclonesTopographic2021}. 
Topographic turbulence is also relevant to long-lived, anticyclonic vortices locked to large-scale topographic bowls in the ocean, such as the Lofoten Basin Eddy and the Mann Eddy in the North Altantic \citep[see][]{kohlGenerationStabilityQuasiPermanent2007,lacasceVorticesBathymetry2024}. 
Since current climate models fall short of resolving turbulent flows in the ocean, employing the topographic turbulence model to investigate phenomena arising from it will facilitate the parameterization of ocean turbulence over bumpy seabeds in these models \citep{hollowayRepresentingTopographicStress1992,radkoGeneralizedTheoryFlow2023,eavesEnergyEnstrophyConstrained2024}.

Theories of topographic turbulence aimed to derive the steady states via variational principles.
For an energy-conserving system, 
\citet{brethertonTwodimensionalTurbulenceTopography1976} conjectured that potential enstrophy is minimized,
leading to a linear relation between the potential vorticity (PV hereafter) and geostrophic streamfunction.
The other approach is to maximize the entropy and is thus related to the equilibrium statistical mechanics \citep{salmonEquilibriumStatisticalMechanics1976}.
The ensembles of PV and streamfunction also satisfy a linear relation. 
According to \citet{carnevaleNonlinearStabilityStatistical1987}, the two solutions are actually consistent with each other, and nonlinearly stable.


Recent numerical experiments show that topographic turbulence shares some common features with flat-bottom two-dimensional turbulence (2DT). 
For instance, condensated states commonly found in 2DT \citep{boffettaTwoDimensionalTurbulence2012} have been identified in topographic turbulence over small-scale topography \citep{zhangSpectralCondensationQuasigeostrophic2023,galletTwodimensionalTurbulenceTopography2024}.
\citet{siegelmanTwodimensionalTurbulenceTopography2023} (SY23 hereafter) showed the coexistence of vortices and a background flow in topographic turbulence, the former of which are prevalent in 2DT \citep{mcwilliamsEmergenceIsolatedCoherent1984,mcwilliamsVorticesTwodimensionalTurbulence1990,benziIntermittencyCoherentStructures1986,benziStatisticalPropertiesTwoDimensional1987,benziSelfsimilarCoherentStructures1988,santangeloGenerationVorticesHighresolution1989}. 
They found that
the mobility of vortices and the uniformity of background PV depend on the energy of the system: on the low-energy branch, the vortices are locked to topographic features, and the background PV is spatially non-uniform; on the high-energy branch, the vortices roam throughout the domain, and the background PV is homogenized \citep{rhinesHomogenizationPotentialVorticity1982}.
These two branches are separated by a critical energy.

SY23 examined the minimum-enstrophy hypothesis of \citet{brethertonTwodimensionalTurbulenceTopography1976} against their numerical simulations, and identified significant discrepancies: the high-energy solutions displayed homogeneous background PV that is not predicted by the minimum-enstrophy hypothesis at the same high energy; the {long-term total} enstrophy far exceeded that of the theoretical prediction for almost all energies.
SY23 attributed these discrepancies to the presence of long-lived vortices: the roaming of vortices mixes the background PV toward homogeneity, and vortex shielding inhibits enstrophy from cascading to smaller scales. 

In this work, we will show that the emergence of vortices might relate to the specific intermediate length scale of the initial fields prescribed by SY23, at which the initial total enstrophy is much higher than the theoretical minimum value and the large margin may serve as the seeds of vortices.
As opposed to fixing an initially energized wavenumber,
we conduct numerical experiments similar to those of SY23 but initialised by monoscale fields sweeping a broad range of wavenumbers, from scales comparable to the domain size to those much smaller than that chosen by SY23.
The purpose here is to investigate how topographic turbulence evolves in response to different scales of initial conditions: 
can the emergence of vortices be inhibited and the minimum-enstrophy state be approached if the initial scale is adequately large; what would occur if the initial scale is smaller than that of SY23?

\section{Framework}

\subsection{Theory}
As in SY23, we consider an unforced, single-layer quasi-geostrophic (QG) flow on a $f$-plane in a doubly-periodic domain ($L\times L$, with $L$ denoting the domain size).
The governing equation of the QG PV $q(x,y,t)$ reads
\refstepcounter{equation}\label{eq:governing_equation}
$$
\frac{\partial q}{\partial t}+\frac{\partial\psi}{\partial x}\frac{\partial q}{\partial y}-\frac{\partial\psi}{\partial y}\frac{\partial q}{\partial x}=D\zeta, \quad q = \zeta+\eta=\nabla^2\psi+\eta, \eqno{(\theequation{\mathit{a},\mathit{b}})}
$$
where $\psi$ and $\zeta$ are the geostrophic streamfunction and relative vorticity, respectively.
The operator $D$ stands for the dissipation of PV that will be specified later.
Effects from bottom topography are encoded in the topographic PV, $\eta(x,~y)=-f_0h_1/h_0$, for a total depth of $h_0+h_1(x,y)$ with small fluctuations $h_1(x,y)$ around a constant average $h_0$, 
 scaled by the local Coriolis frequency $f_0$ (see SY23).

In the absence of dissipation, the system (\ref{eq:governing_equation}) conserves the kinetic energy and total enstrophy, defined respectively as
\refstepcounter{equation}
$$
E = \frac{1}{2}\left<|\nabla\psi|^2\right>= \frac{1}{2L^2}\iint|\nabla\psi|^2dxdy \quad \mbox{and} \quad Q = \frac{1}{2}\left<q^2\right>=\frac{1}{2L^2}\iint q^2dxdy. \eqno{(\theequation{\mathit{a},\mathit{b}})}
$$
Here $\left<\cdot\right>$ represents an domain-average operation.
When the added dissipation in the governing equation (\ref{eq:governing_equation}) operates at high wavenumbers, the energy is approximately conserved, while the enstrophy drops by a large amount as a result of forward cascade.
This motivated \citet{brethertonTwodimensionalTurbulenceTopography1976} to predict that an energy-conserving system would evolve into a minimum-enstrophy state. 
Such a state can be obtained by performing variational calculation of the functional $\mathcal{L}=Q+\mu E$, with $\mu$ denoting the Lagrange multiplier. The resulting Euler-Lagrange equation,
\begin{equation}\label{eq:Euler-Lagrange_equation}
    q_\ast=\nabla^2\psi_\ast+\eta=\mu\psi_\ast,
\end{equation}
dictates a linear relation between $q_\ast$ and $\psi_\ast$.
Solutions to (\ref{eq:Euler-Lagrange_equation}) are inviscid steady solutions to the QG PV equation (\ref{eq:governing_equation}).

\begin{figure}
    \centering
    \includegraphics[width=0.4\textwidth]{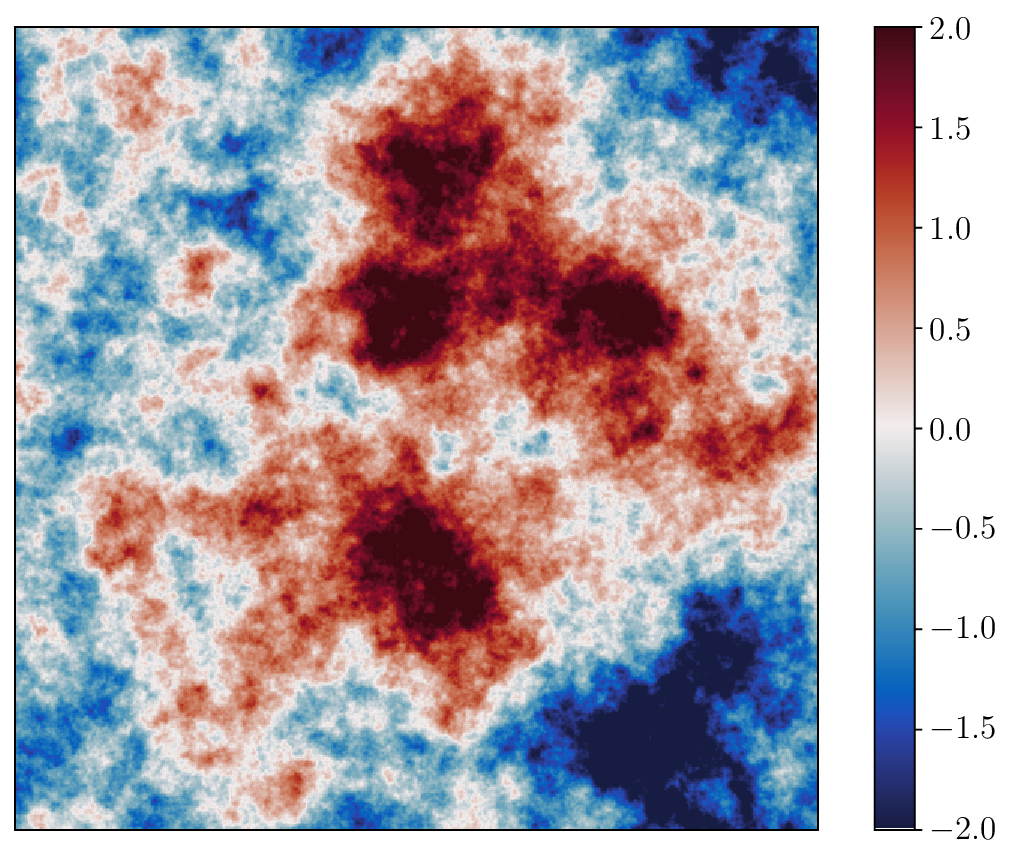}
    \caption{Topographic potential vorticity $\eta(x,y) (\mathrm{unit:}~ 10^{-6}~\mathrm{s}^{-1})$ used throughout this work.}
    \label{fig:topography}
\end{figure}

To solve the equation (\ref{eq:Euler-Lagrange_equation}),
the topographic PV can be Fourier-transformed into
\begin{equation}\label{eq:topography_fourier_series}
\eta(x,~y) = \sum_{\bk}\eta_{\bk}\rme^{\rmi\bk\cdot\boldsymbol{x}},
\end{equation}
where $\bk$ represents the wavenumber vector.
As in SY23, the topographic PV is chosen to satisfy a power spectrum of $k^{-2}$ ($k=|\bk|$), by which the Fourier amplitudes in the expression (\ref{eq:topography_fourier_series}) shall take the form
\begin{equation}\label{eq:TopoPVFourierComp}
    \eta_{\bk}=\alpha\rme^{\rmi\phi_{\bk}}k^{-3/2}.
\end{equation}
The normalization parameter $\alpha$ is adjusted to produce a specific value of the root mean square of the topographic PV, defined as $\eta_{rms}=\sqrt{\sum_{\bk}|\eta_{\bk}|^2}$.
Following SY23, we prescribe $\eta_{rms} = 10^{-6}~\mathrm{s}^{-1}$. Meanwhile,
the random phase $\phi_{\bk}$ in (\ref{eq:TopoPVFourierComp}) serves to produce a realization of $\eta$.
Throughout this work, the topography remains fixed, and is shown in figure \ref{fig:topography}.

With the prescribed topography, the solution to the Euler-Lagrange equation (\ref{eq:Euler-Lagrange_equation}) can be obtained in spectral space as
\begin{equation}\label{eq:Euler-Lagrange_solution}
\psi_\ast(x,y,\mu)=
\sum_{\bk}\frac{\eta_{\bk}\rme^{\rmi\bk\cdot\boldsymbol{x}}}{\mu+k^2}.
\end{equation}
The corresponding energy and minimum enstrophy are then
\refstepcounter{equation}\label{eq:Euler-Lagrange_energy_enstrophy}
$$
E=\frac{1}{2}\sum_{\bk}\frac{k^2|\eta_{\bk}|^2}{\left(\mu+k^2\right)^2} \quad \mbox{and} \quad Q_{min}=\frac{1}{2}\sum_{\bk}\frac{\mu^2|\eta_{\bk}|^2}{\left(\mu+k^2\right)^2}, \eqno{(\theequation{\mathit{a},\mathit{b}})}
$$
respectively.
With a given energy level $E$ for an energy-conserving system, the Lagrange multiplier $\mu$ and minimum enstrophy $Q_{min}$ can be sought through the above relations.

\begin{figure}
    \centering
    \includegraphics[width=1\textwidth]{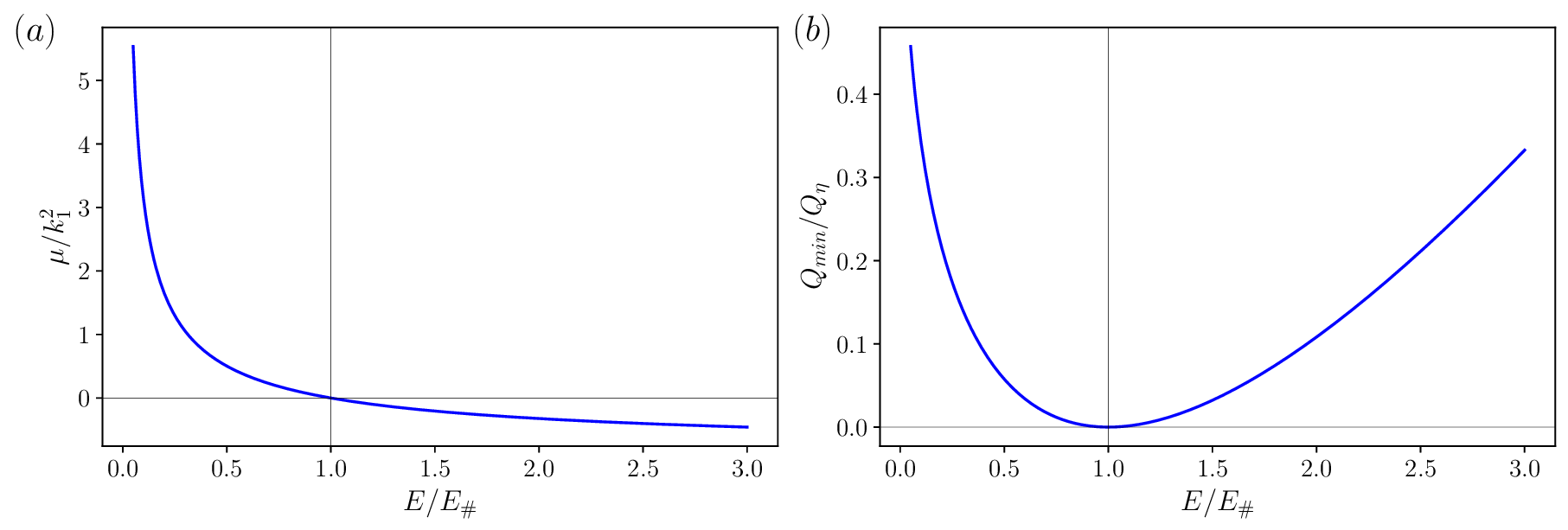}
    \caption{Minimum-enstrophy solutions as in SY23 (see their Figures 2(a)--(b)).}
    \label{fig:minimum_enstrophy_solution}
\end{figure}

Figure \ref{fig:minimum_enstrophy_solution}$(a,b)$ show the resulting $\mu$ and $Q_{min}$ as functions of $E$, respectively, scaled by $k_1^2$ ($k_1=2\pi/L$ represents the fundamental wavenumber) and $Q_{\eta}=\eta_{rms}^2/2$. 
In addition, the energy has been scaled by the critical energy level $E_\#$, corresponding to the particular case of $\mu=0$ and thus of homogenized PV ($q_\#=0$). At this critical energy,
the minimum enstrophy vanishes ($Q_{min}=0$; see figure \ref{fig:minimum_enstrophy_solution}$(b)$). 
The critical energy $E_\#$ thus separates the low- and high-energy branches, which correspond to positive and negative values of $\mu$, respectively. These theoretical solutions are identical to those shown in SY23 (see their figure~2$(a,b)$), who further found that $E_\#$ is a good separator of high- and low-energy branches of their numerical experiments.

\subsection{Numerical setups}
We solve the QG PV equation (\ref{eq:governing_equation}) using the open-source pseudo-spectral package \textsf{GeophysicalFlows.jl} running on GPUs \citep{constantinouGeophysicalFlowsJlSolvers2021}.
Time is stepped forward by a fourth-order Rounge-Kutta scheme.
The dissipation term ($D\zeta$) in (\ref{eq:governing_equation}) is implemented via a spectral filter of relative vorticity, which is applied to high wavenumbers at the end of each time step.
We choose exactly the same parameter setups as in SY23: 
the domain size is $10^6$ meters; 
the resolution is $1024\times1024$, for which the spectral filter is applied to wavenumbers higher than the cutoff wavenumber $k_{\mathrm{cutoff}}(=2/3\times512k_1\approx341k_1)$;
the integration time is $1.5\times10^9$ seconds (around $47.53$ years); 
time step is $1500$ seconds, which is reduced further to $500$ seconds for computationally demanding runs.

\begin{figure}
    \centering
    \includegraphics[width=0.75\textwidth]{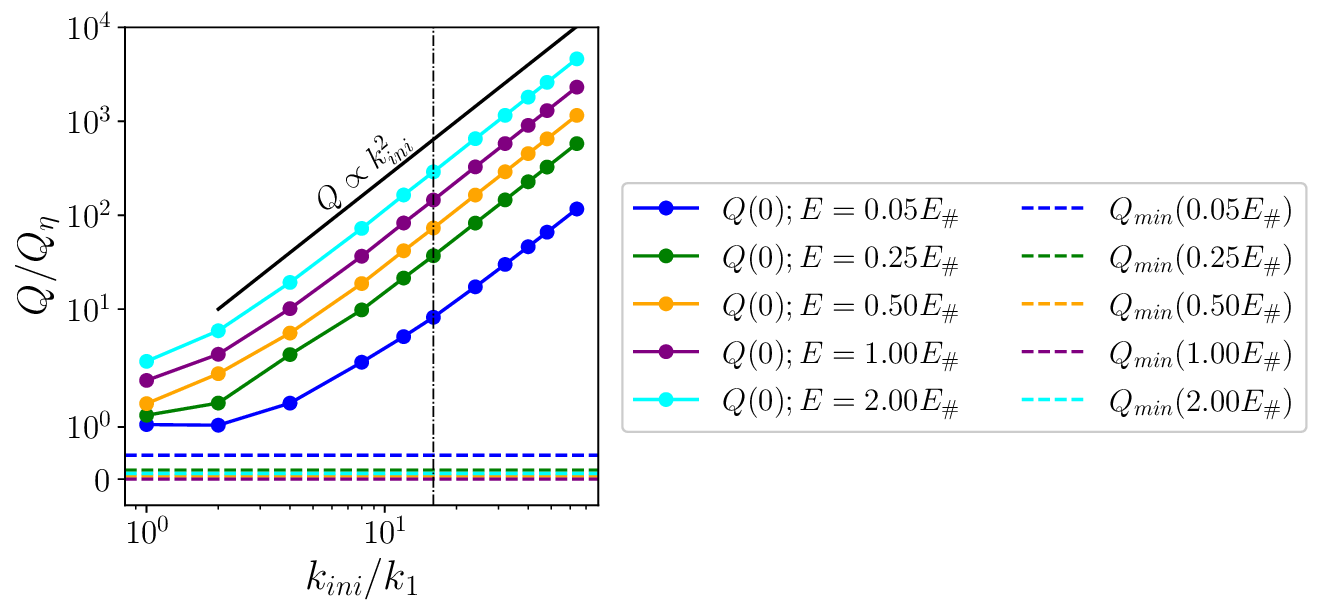}
    \caption{Variations of initial enstrophy $Q(0)$ with the initial  wavenumber $k_{ini}$ for different energies, compared with the minimum enstrophy $Q_{min}$.
    The vertical axis is in symmetric log scale, which is linear within $[0,1]$ and logarithmic outside it.}
    \label{fig:initial_conditions}
\end{figure}

We initialise the computations by a series of random monoscale relative vorticity fields with a wide range of initial enstrophy, which is achieved by different combinations of the wavenumber and energy level.
For all cases, the Fourier components of the initial relative vorticity $\zeta(x,y,0)$ are chosen randomly within the limited bandwidth $[k_{ini}-0.5k_1, k_{ini}+0.5k_1]$, where $k_{ini}$ is the wavenumber of the initial monoscale field.
Note that the selected $k_{ini}$ should be much smaller than the cutoff wavenumber $k_{\mathrm{cutoff}}\approx341k_1$ to avoid direct influence of the spectral filter on the initial condition and to remove enstrophy aggregated at wavenumbers higher than $k_{\mathrm{cutoff}}$.
In practice, we choose a series of 
\begin{equation}\label{eq:wavenumbers}
    k_{ini} \in \{1, 2, 4, 8, 12, 16, 24, 32, 48, 64\}k_1,
\end{equation}
corresponding to large-to-small initial monoscales, and 
\begin{equation}
    E \in \{0.05, 0.15, 0.25, 0.35, 0.50, 0.75, 1.00, 2.00\}E_\#,
\end{equation}
corresponding to low-to-high energy levels. 
Note that SY23 prescribed the initial monoscale at approximately $16k_1$, which is intermediate in our selected range of wavenumbers (\ref{eq:wavenumbers}).
The initial enstrophy for different combinations of $k_{ini}$ and $E$ is shown in figure \ref{fig:initial_conditions}, estimated as $Q(0) \approx k_{ini}^2E + Q_{\eta}$ (the cross term between the initial flow and topography is found to be negligible in magnitude). That is,
with a prescribed energy level, $Q(0)$ increases at a rate of approximately $k_{ini}^2$, which means that larger amounts of initial enstrophy are contained at smaller scales.
As shown in figure \ref{fig:initial_conditions}, the initial enstrophy exceeds the theoretical minimum enstrophy for all cases.
However, the former is adequately close to the latter in the small-$k_{ini}$ regime, as opposed to the case of intermediate wavenumber (highlighted by the vertical dotted-dashed line in figure \ref{fig:initial_conditions}) considered by SY23. Owing to the initial/minimum enstrophy proximity,
small-$k_{ini}$ runs may readily converge to the minimum-enstrophy states. 

\section{Results}
In this section, we analyse the solutions of our numerical experiments.
Particular attention is paid to the dependence of the long-term state on the initial wavenumber $k_{ini}$.
Results of $k_{ini}=16k_1$, consistent with those in SY23, are highlighted in red color upon comparison with other wavenumber cases.

\subsection{Time evolution of energy and enstrophy}
\begin{figure}
    \centering
    \includegraphics[width=0.8\textwidth]{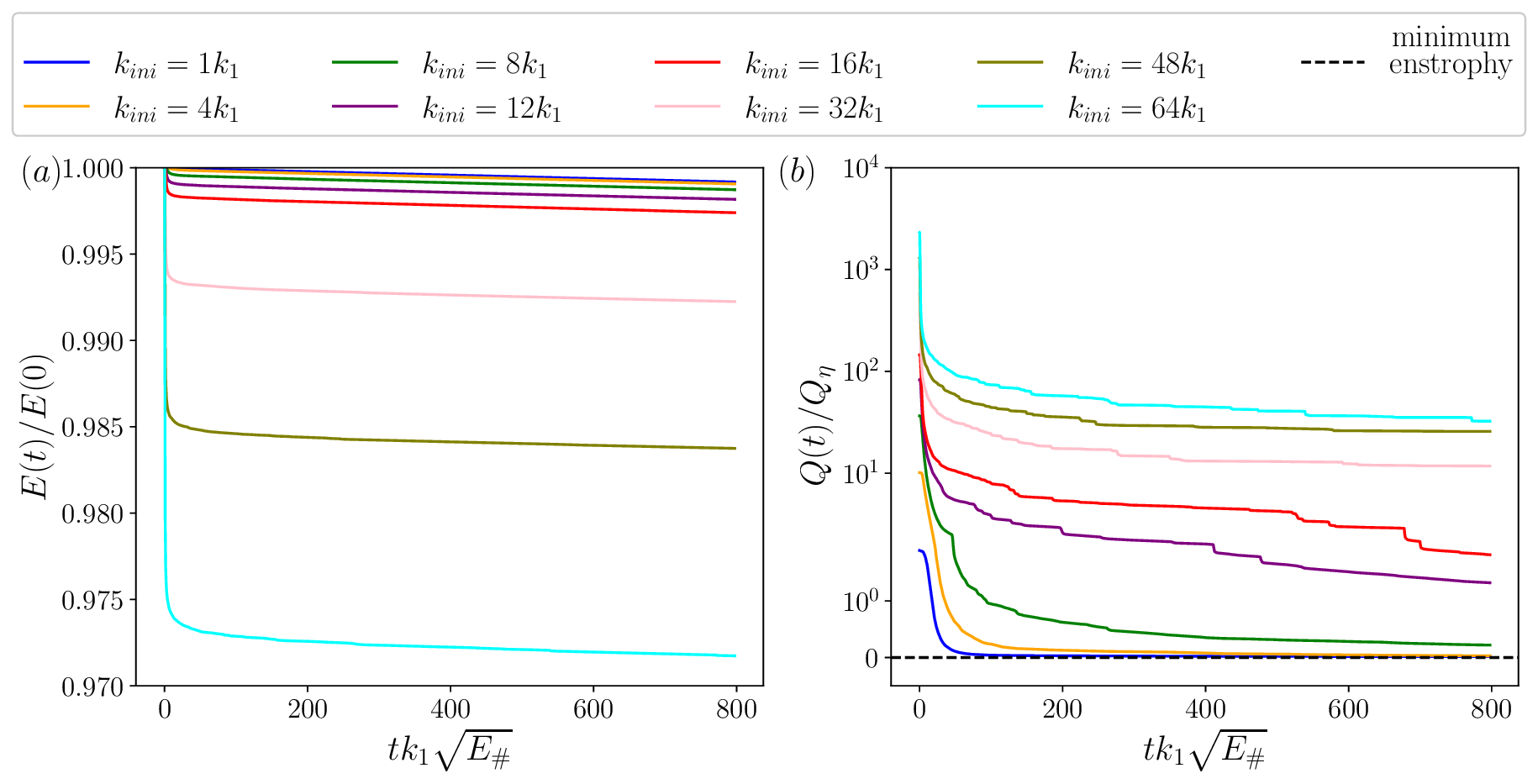}
    \caption{Time evolution of energy ($a$) and enstrophy ($b$) for the $E_\#$ runs. The vertical axis of $(b)$ is in symmetric log scale.}
    \label{fig:1.00E_energy_enstrophy}
\end{figure}

The time evolution of energy and enstrophy is shown in figure \ref{fig:1.00E_energy_enstrophy} for simulations of the critical energy $E/E_\#=1$
(results of other energies exhibit similar behaviours and are not shown).
As shown in figure \ref{fig:1.00E_energy_enstrophy}($a$), the energy loss during run time is larger for larger $k_{ini}$, because the corresponding initial spectral position of energy injection is closer to the filtering region at high wavenumbers.
However, the energy loss at the final time (47.53 years) amounts to a negligible fraction of the initial energy, within $0.3\%$ for $k_{ini}\le 16k_1$ and around $3\%$ for $k_{ini}=64k_1$ (see first column of numbers in table \ref{tab:energy_enstrophy_at_final_time}).
Therefore, energy is nearly conserved for all runs, which is the necessary condition for examining the minimum-enstrophy hypothesis.

\begin{table*}
  \begin{center}
  \begin{tabular}{l|ccccc}
  \hline
      \diagbox{$k_{ini}$}{$E$}  & $0.05E_\#$   &   $0.25E_\#$ & $0.50E_\#$ & $E_\#$ & $2E_\#$ \\ \hline
       theory     & 1.000; 0.457 & 1.000; 0.174 & 1.000; 0.057 & 1.000; 0.000 & 1.000; 0.108 \\[2pt]
       $1{k}_1$   & 0.997; 0.465 & 0.999; 0.177 & 0.999; 0.061 & 0.999; 0.017 & 0.999; 0.193 \\[2pt]
       $2{k}_1$   & 0.997; 0.472 & 0.999; 0.184 & 0.999; 0.064 & 0.999; 0.006 & 0.999; 0.299 \\[2pt]
       $4{k}_1$   & 0.997; 0.475 & 0.998; 0.181 & 0.999; 0.092 & 0.999; 0.030 & 0.999; 2.105 \\[2pt]
       $8{k}_1$   & 0.997; 0.468 & 0.998; 0.224 & 0.998; 0.323 & 0.999; 0.220 & 0.999; 2.420 \\[2pt]
       $12{k}_1$  & 0.996; 0.480 & 0.998; 0.605 & 0.998; 1.173 & 0.998; 1.321 & 0.998; 5.238 \\[2pt]
       $16{k}_1$  & 0.996; 0.582 & 0.997; 1.305 & 0.997; 2.081 & 0.997; 1.816 & 0.998; 8.734 \\[2pt]
       $32{k}_1$  & 0.990; 1.170 & 0.992; 3.835 & 0.992; 5.766 & 0.992; 11.76 & 0.992; 20.98 \\[2pt]
       $48{k}_1$  & 0.981; 2.022 & 0.983; 6.945 & 0.983; 9.439 & 0.984; 25.73 & 0.984; 36.76 \\[2pt]
       $64{k}_1$  & 0.969; 3.115 & 0.971; 11.12 & 0.971; 16.50 & 0.972; 32.41 & 0.972; 63.38 \\\hline
  \end{tabular}
  \caption{Energy $E(47.53~\mbox{years})/E(0)$ (first column of numbers) and enstrophy $Q(47.53~\mbox{years})/Q_\eta$ (second column of numbers) for runs with different $k_{ini}$ and $E$, compared against the theoretical predictions (first row).}
  \label{tab:energy_enstrophy_at_final_time}
  \end{center}
\end{table*}

By contrast, the enstrophy decreases drastically during run time, as shown in figure \ref{fig:1.00E_energy_enstrophy}($b$).
Consistent with the findings of SY23, there is an initial fast drop followed by a slow decrease.
Generally, for a fixed energy level, larger-$k_{ini}$ runs, or equivalently, runs with higher initial enstrophy (see figure \ref{fig:initial_conditions}), maintain higher enstrophy during run time. Runs with small $k_{ini}$ are intriguing: the enstrophy approaches the minimum value $Q_{min}$ during run time.
At the final time (see second column of numbers in table \ref{tab:energy_enstrophy_at_final_time}), the enstrophy $Q$ of small-$k_{ini}$ ($1k_1$, $2k_1$, $4k_1$) runs prescribed with low energy ($E/E_\# \le 1$) is extremely close to the corresponding minimum value $Q_{min}$.
However, for the high energy case of $2E_\#$, the final enstrophy of the $1k_1$ run is almost twice the minimum value.
For the intermediate case of $k_{ini}=16k_1$ considered by SY23 and cases with larger $k_{ini}$, the final enstrophy is much larger than the minimum value, except for the lowest-energy runs ($0.05E_\#$).
Thus, the minimum-enstrophy states can actually be approached by runs with large-scale initial conditions, especially by those with low energy ($E/E_\#\le 1$). These findings contrast markedly with the  conclusions of SY23.

\subsection{Long-term states}
Next we inspect the long-term snapshots in runs with typical energy levels at the final time (47.53 years), and compare those with the minimum-enstrophy states $\zeta_\ast$ and $q_\ast$ obtained theoretically.
The corresponding animations, showing the time evolution to the long-term states, can be found in the supplementary materials.

\begin{figure}
    \centering
    \includegraphics[width=0.7\textwidth]{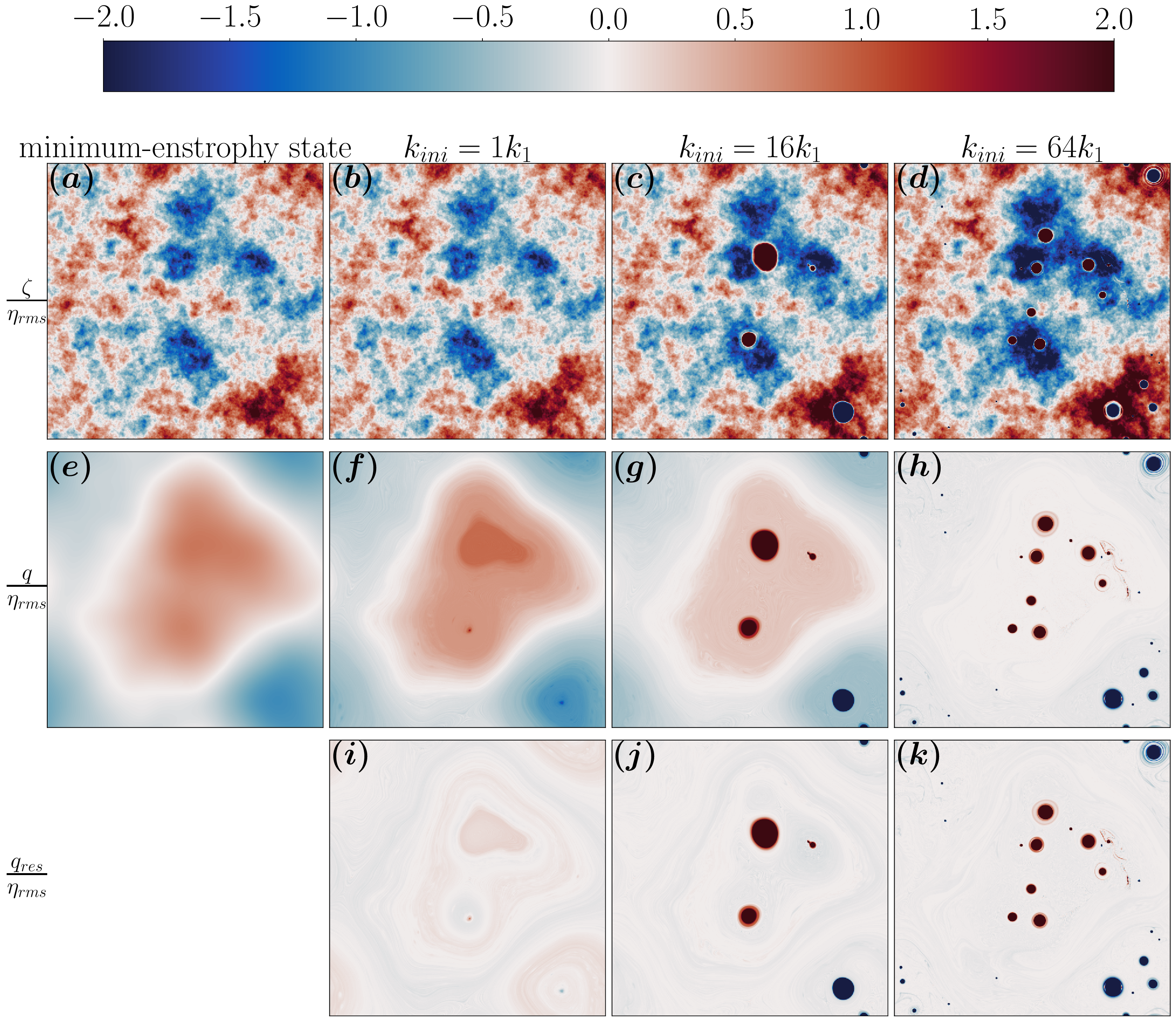}
    \caption{Long-term snapshots of $0.25E_\#$ runs for $k_{ini}=[1, 16, 64]k_1$ (second-fourth columns) compared with the minimum enstrophy state at the same energy level (first column): first row, $\zeta/\eta_{rms}$; second row, $q/\eta_{rms}$; third row, $q_{res}/\eta_{rms}$.}
    \label{fig:final_snapshots_0.25E}
\end{figure}

Figure \ref{fig:final_snapshots_0.25E} compares the snapshots of relative vorticity (first row) and PV (second row) from three runs with different $k_{ini}$ against the minimum-enstrophy state at the low energy of $E=0.25E_\#$.
The state of the smallest $k_{ini}=1k_1$ run (figure \ref{fig:final_snapshots_0.25E}($b,f$)) resembles the minimum-enstrophy state (figure \ref{fig:final_snapshots_0.25E}($a,e$)), displaying a non-uniform PV field shaped by the low-pass-filtered topography throughout the domain.
This resemblance is consistent with the proximity of the corresponding enstrophy (see table \ref{tab:energy_enstrophy_at_final_time}), and further demonstrates that the minimum-enstrophy state can actually be approached with very large-scale initial conditions.
However, although no strong vortices emerge, there exist some weak extrema of PV (see figure \ref{fig:final_snapshots_0.25E}$(f)$) at the topographic depressions and elevations (see figure \ref{fig:topography} for topography).
As $k_{ini}$ increases to $16k_1$, strong vortices emerge at these locations, with cyclones and anticyclones locked to topographic elevations and depressions, respectively (figure \ref{fig:final_snapshots_0.25E}($g$)).
The background relative vorticity becomes stronger (compare figure \ref{fig:final_snapshots_0.25E}($c$) with ($b$)), leading to the weaker background PV (compare figure \ref{fig:final_snapshots_0.25E}($g$) with ($f$)).
As $k_{ini}$ increases further to $64k_1$, more vortices nucleate.
Notably, the background relative vorticity strengthens further and the background PV is nearly homogenized (figure \ref{fig:final_snapshots_0.25E}($d,h$)). A larger initial wavenumber provides more seeds of vortices; more vortices nucleate afterwards and mix the background PV more efficiently to homogeneity.
Note that the vortices in the $64k_1$ run remain locked to fixed locations but exhibit higher mobility compared to those in the $16k_1$ run (see supplementary movies of the corresponding runs).
In summary, at a low energy level, the resemblance with the minimum-enstrophy state and the near homogenization of background PV are observed in simulations with two limiting initial length scales, with an emergent transitional state as the initial length scale varies in between.
These phenomena cannot be observed in low-energy simulations with a single intermediate wavenumber of $16k_1$ considered by SY23.

\begin{figure}
    \centering
    \includegraphics[width=0.85\textwidth]{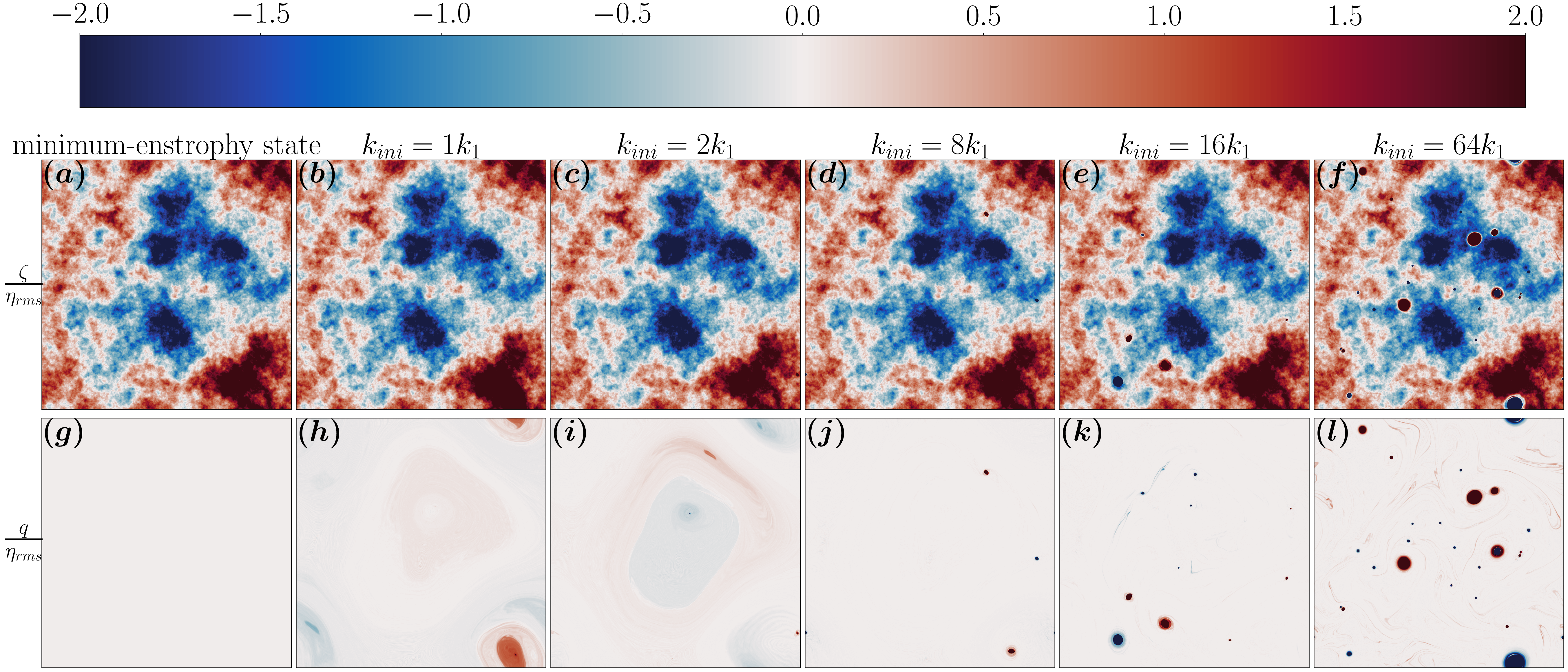}
    \caption{Long-term snapshots of $E_\#$ runs for $k_{ini}=[1, 2, 8, 16, 64]k_1$ (second-sixth columns) compared with the minimum enstrophy state at the same energy level (first column): first row, $\zeta/\eta_{rms}$; second row, $q/\eta_{rms}$.}
    \label{fig:final_snapshots_1.00E}
\end{figure}

Now we turn to the runs with the critical energy level of $E_\#$, as shown in figure \ref{fig:final_snapshots_1.00E}.
Once again, the solutions of the small-$k_{ini}$ runs are similar to the minimum-enstrophy state, with nearly homogenized  PV (see figure \ref{fig:final_snapshots_1.00E}$(h,i)$). This demonstrates that the minimum-enstrophy state can be approached, together with the analogous enstrophy produced numerically and theoretically (table \ref{tab:energy_enstrophy_at_final_time}).
As $k_{ini}$ increases, more vortices nucleate and roam throughout the domain.
The background PV is then completely homogenized.

\begin{figure}
    \centering
    \includegraphics[width=0.85\textwidth]{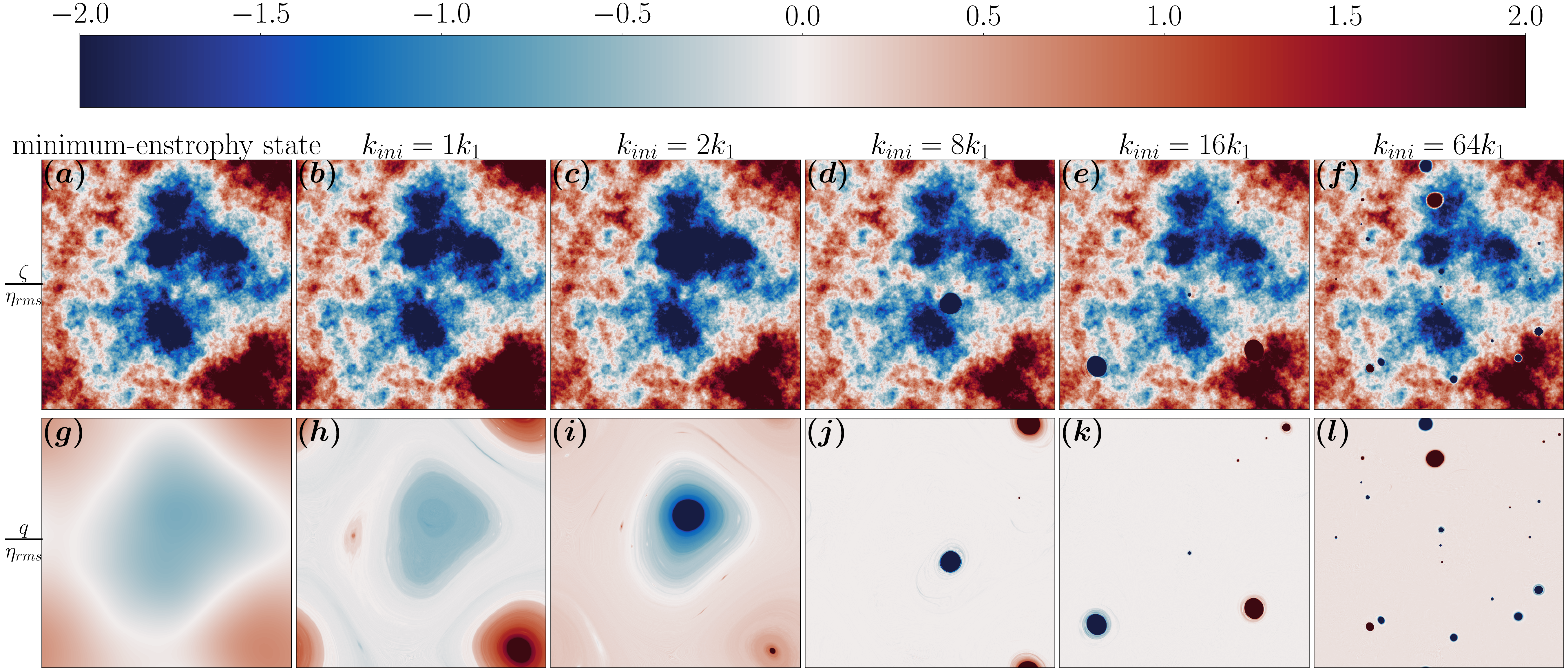}
    \caption{Same as figure \ref{fig:final_snapshots_1.00E} but for $2E_\#$ runs.}
    \label{fig:final_snapshots_2.00E}
\end{figure}

A parallel comparison is made further among simulations prescribed with the high energy level of $2E_\#$, shown in figure \ref{fig:final_snapshots_2.00E}.
As discussed earlier, the final-state enstrophy does not approach the minimum value, even for the case of the smallest initial wavenumber $1k_1$ (table \ref{tab:energy_enstrophy_at_final_time}).
However, the solution of the $1k_1$ run is still similar to the minimum enstrophy state to some extent (figure \ref{fig:final_snapshots_2.00E}$(g,h)$), both displaying non-uniform PV fields imprinted by the low-pass-filtered topography.
As shown in figure \ref{fig:final_snapshots_2.00E}$(h,i)$ for runs of $1k_1$ and $2k_1$, a cyclone and a anticyclone are locked respectively to the topographic depression and elevation, exactly opposing the relation between vortex polarity and depth in the low-energy runs.
These gigantic vortices inhibit enstrophy from further decaying toward the theoretical minimum value, and result likely from some subtle, incomplete adjustments of topographic turbulence suggested by \citet{lacasceVorticesBathymetry2024}.
Toward the case of $8k_1$, the vortices become smaller in size and more localised; they orbit around the large-scale features of topography (see the corresponding supplementary movie).
The background PV is essentially homogenized.
As $k_{ini}$ increases further, more vortices emerge and tend to roam throughout the domain.
Note that, although the background PV of the $64k_1$ run is homogenized, it is not identically zero.

\subsection{Empirical background flow}

\begin{figure}
    \centering
    \includegraphics[width=\textwidth]{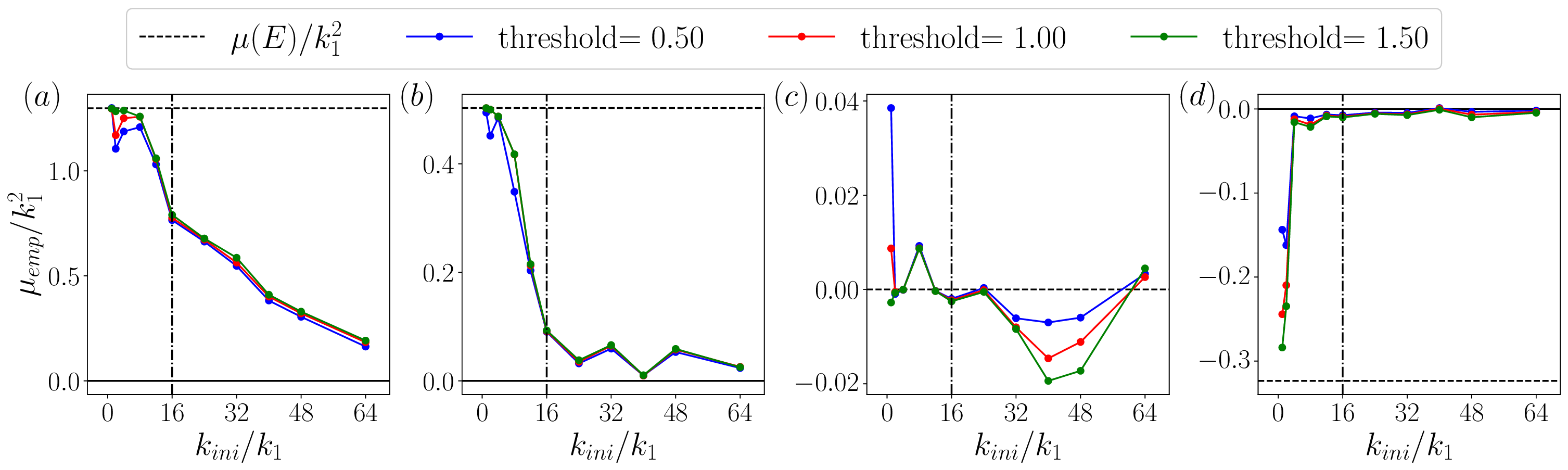}
    \caption{Variations of the empirical slope $\mu_{emp}$ between $q$ and $\psi$ with the initial wavenumber $k_{ini}/k_1$ at four energy levels: $(a)$ $E=0.25E_\#$; $(b)$ $E=0.50E_\#$; $(c)$ $E=E_\#$; $(d)$ $E=2E_\#$. Horizontal dashed lines represent the theoretical Lagrange multiplier based on the minimum-enstrophy principle at the same energy levels. Vertical dotted-dashed lines highlight the wavenumber of $k_{init}/k_1 = 16$ considered in SY23.}
    \label{fig:multipliers}
\end{figure}

As in SY23, we extract the background flow from the data via linear fitting between $q$ and $\psi$.
The vortices in the flow field are outliers and removed by a threshold. 
The points with $|q|/\eta_{rms}$ below the threshold are retained to extract the background state.
Then applying least-squares fitting to $q$ and $\psi$ of the remaining points yields an approximate linear relation,
\begin{equation}\label{eq:emp_linear}
    q\approx \mu_{emp}\psi.
\end{equation}
When extracting the empirical slope $\mu_{emp}$ for each time instant, we found that its value does not vary significantly in time after an initial adjustment phase. 
The empirical slope $\mu_{emp}$ should depend on the energy $E$, as illustrated by SY23, and the initial wavenumber $k_{ini}$ based on our experiments.
The results of $\mu_{emp}$ are shown in figure \ref{fig:multipliers}.
We choose three thresholds for extracting $\mu_{emp}$, and find only quantitative differences.
The runs with the critical energy  $E_\#$ show that $\mu_{emp}$ is always close to zero, indicating near homogenization of PV.
For small-$k_{ini}$ runs, the empirical slope $\mu_{emp}$ is close to the theoretical $\mu(E)$ (horizontal dashed line) shown in figure \ref{fig:minimum_enstrophy_solution}$(a)$.
Even at the high energy level of $2E_\#$, a finite and negative $\mu_{emp}$ is consistent with the theoretical $\mu(2E_\#)$.
This is another justification of the existence of the minimum enstrophy state approachable by numerical simulations but not obtained by SY23.
As the initial wavenumber increases, the magnitude of $\mu_{emp}$ tends to zero, implying a transition from a non-uniform background PV field imprinted by the low-pass-filtered topography to PV homogenization, consistent with the preceding observations from the long-term snapshots. The numerical solutions of SY23 (see vertical dotted-dashed lines) might capture a confined range of the transition regime, particularly in the low-energy cases.

As in SY23, the empirical slope $\mu_{emp}$ is used to define a ``residual'' PV field, $q_{res} = q-\mu_{emp}\psi$,
which is shown in figure \ref{fig:final_snapshots_0.25E}$(i,j,k)$.
Evidently, the background flow is sufficiently removed, thus confirming the $q$-$\psi$ linear relation (\ref{eq:emp_linear}).

\begin{figure}
    \centering
    \includegraphics[width=0.7\textwidth]{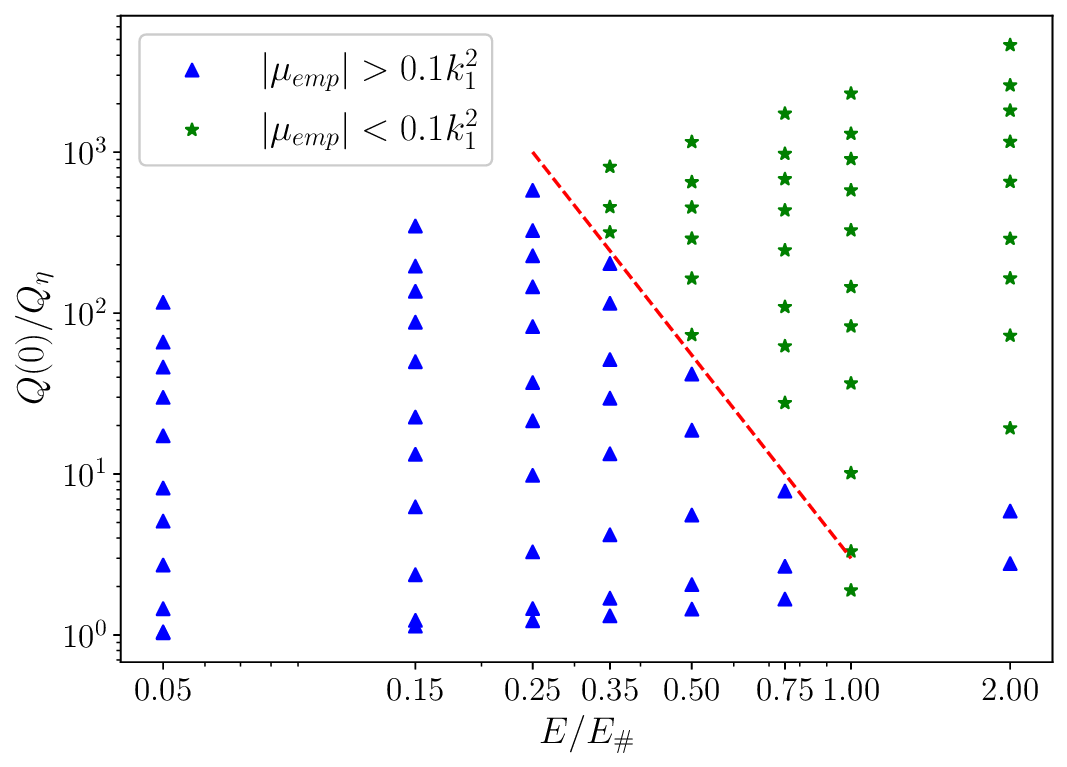}
    \caption{Phase diagram of the runs in the parametric space of the energy level $E/E_\#$ and the initial enstrophy $Q(0)/Q_{\eta}$ (both in log scales). Markers $\textcolor{blue}{\blacktriangle}$ and $\textcolor{OliveGreen}{\star}$ denote the runs with $\mu_{emp}$ larger and smaller than $0.1k_1^2$ respectively. The red dashed line roughly delineates the regimes of non-uniform and homogeneous background PV in the low-energy region $E<E_\#$.}
    \label{fig:phase_diagram}
\end{figure}

As shown in figure \ref{fig:initial_conditions}, the initial length scale directly modulates the initial enstrophy.
Thus, the dependence of the long-term states on the former can be regarded as to that on the latter.
Figure \ref{fig:phase_diagram} depicts a phase diagram of all runs in the parametric space of the energy level $E/E_\#$ and the initial enstrophy $Q(0)/Q_\eta$.
Two regimes of the long-term states, namely, non-uniform and homogeneous background PV, are delineated by the empirical slope $\mu_{emp}$ in comparison against a small threshold of $0.1k_1^2$.
The two regimes are roughly separated by a straight line on the low-energy branch $E<E_\#$.
We can observe that lower-energy runs need higher initial enstrophy to approach homogeneous background PV.
This can be understood by the fact that the minimum-enstrophy PV with lower energy is more non-uniform and requires stronger mixing (promoted by higher enstrophy) to approach homogenization.

\section{Conclusion}
In this work, we study the states of unforced, weakly-decaying two-dimensional turbulence above random rough topography, initialised by a series of monoscale fields covering a broad range of length scales and energy levels.
Our observations complement those of \citet{siegelmanTwodimensionalTurbulenceTopography2023}, who took a single, intermediate monoscale for initialisation into account.
Some of the phenomenological descriptions of topographic turbulnece from \citet{siegelmanTwodimensionalTurbulenceTopography2023} shall be updated.
The minimum-enstrophy solutions of \citet{brethertonTwodimensionalTurbulenceTopography1976} can be approached by initial fields whose length scales are comparable with the domain size, especially when the energy is no larger than $E_\#$.
As the initial length scale decreases, 
the higher initial enstrophy provides more seeds for vortex nucleation.
More emerging vortices shield enstrophy from cascading toward high wavenumbers, and efficiently mix the background PV.
A non-uniform background PV field imprinted by large-scale topography, at low or high energy, is thus weakened and tends to homogenization.
Our results highlight the sensitivity of topographic turbulence to the length scales of initial conditions. Given that the depth-invariant mode of geostrophic turbulence responsible for driving mass and heat transports was constantly found to be energized across a range of wavenumbers \citep{ChenTwolayerQGTurbulence23}, this work calls for ongoing efforts for refining parameterisations of ocean turbulence over bumpy seafloor by taking the energization length scales into account.

Our work has confined the initial conditions of topographic turbulence to monoscales.
How the turbulence responds to multi-scale initial conditions, such as a power-law field, is yet to be addressed.
For simplicity, if one considers a superposition of two monoscale fields with identical energy and enstrophy to those of a single monoscale field, the energy and enstrophy must be redistributed between the two scales, leading to one scale larger than and the other smaller than the single monoscale.
According to our monoscale experiments, a large-scale initial field tends to the minimum-enstrophy state whose PV is typically non-uniform, whereas a small-scale counterpart tends to generate vortices that mix the background PV to homogenization.
Thus, there could be a competition among different scales, in term of bringing the background PV to non-uniformity or homogenization.
Which effect is dominant may depend on the assigned energy to the two monoscales, among other processes. These possibilities open a future research avenue in topographic turbulence.


\backsection[Acknowledgements]{This work is supported by the Research Grants Council (RGC) of Hong Kong under awards Early Career Scheme 26307720 and General Research Fund 16305321, and by the Center for Ocean Research (CORE), a joint research center between Laoshan Laboratory and HKUST.
The authors are grateful to the three anonymous reviewers for their constructive comments which improved this paper a lot.
}


\backsection[Declaration of interests]{The authors report no conflict of interest.}






\bibliographystyle{jfm}
\bibliography{jfm}

\begin{thebibliography}{23}
\expandafter\ifx\csname natexlab\endcsname\relax\def\natexlab#1{#1}\fi
\def\au#1{#1} \def\ed#1{#1} \def\yr#1{#1}\def\at#1{#1}\def\jt#1{\textit{#1}} \def\bt#1{#1}\def\bvol#1{\textbf{#1}} \def\vol#1{#1} \def\pg#1{#1} \def\publ#1{#1}\def\arxiv#1{#1}\def\org#1{#1}\def\st#1{\textit{#1}}

\bibitem[Benzi {\em et~al.\/}(1986)Benzi, Paladin, Patarnello, Santangelo \& Vulpiani]{benziIntermittencyCoherentStructures1986}
{\sc \au{Benzi, R.}, \au{Paladin, G.}, \au{Patarnello, S.}, \au{Santangelo, P.} \& \au{Vulpiani, A.}} \yr{1986}  \at{Intermittency and coherent structures in two-dimensional turbulence}.  \jt{J. Phys. A: Math. Gen.}  \bvol{19}~(18),  \pg{3771}.

\bibitem[Benzi {\em et~al.\/}(1987)Benzi, Patarnello \& Santangelo]{benziStatisticalPropertiesTwoDimensional1987}
{\sc \au{Benzi, R.}, \au{Patarnello, S.} \& \au{Santangelo, P.}} \yr{1987}  \at{On the {{Statistical Properties}} of {{Two-Dimensional Decaying Turbulence}}}.  \jt{EPL}  \bvol{3}~(7),  \pg{811}.

\bibitem[Benzi {\em et~al.\/}(1988)Benzi, Patarnello \& Santangelo]{benziSelfsimilarCoherentStructures1988}
{\sc \au{Benzi, R.}, \au{Patarnello, S.} \& \au{Santangelo, P.}} \yr{1988}  \at{Self-similar coherent structures in two-dimensional decaying turbulence}.  \jt{J. Phys. A: Math. Gen.}  \bvol{21}~(5),  \pg{1221}.

\bibitem[Boffetta \& Ecke(2012)]{boffettaTwoDimensionalTurbulence2012}
{\sc \au{Boffetta, G.} \& \au{Ecke, R.~E.}} \yr{2012}  \at{Two-{{Dimensional Turbulence}}}.  \jt{Annu. Rev. Fluid Mech.}  \bvol{44}~(1),  \pg{427--451}.

\bibitem[Bretherton \& Haidvogel(1976)]{brethertonTwodimensionalTurbulenceTopography1976}
{\sc \au{Bretherton, F.~P.} \& \au{Haidvogel, D.~B.}} \yr{1976}  \at{Two-dimensional turbulence above topography}.  \jt{J. Fluid Mech.}  \bvol{78}~(1),  \pg{129--154}.

\bibitem[Carnevale \& Frederiksen(1987)]{carnevaleNonlinearStabilityStatistical1987}
{\sc \au{Carnevale, G.~F.} \& \au{Frederiksen, J.~S.}} \yr{1987}  \at{Nonlinear stability and statistical mechanics of flow over topography}.  \jt{J. Fluid Mech.}  \bvol{175}~(-1),  \pg{157}.

\bibitem[Chen(2023)]{ChenTwolayerQGTurbulence23}
{\sc \au{Chen, S.-N.}} \yr{2023}  \at{Revisiting the baroclinic eddy scalings in two-layer, quasigeostrophic turbulence: Effects of partial barotropization}.  \jt{J. Phys. Oceanogr.}  \bvol{53}~(3),  \pg{891 -- 913}.

\bibitem[Constantinou {\em et~al.\/}(2021)Constantinou, Wagner, Siegelman, Pearson \& Pal{\'o}czy]{constantinouGeophysicalFlowsJlSolvers2021}
{\sc \au{Constantinou, Navid~C.}, \au{Wagner, Gregory~LeClaire}, \au{Siegelman, Lia}, \au{Pearson, Brodie~C.} \& \au{Pal{\'o}czy, Andr{\'e}}} \yr{2021}  \at{{{GeophysicalFlows}}.jl: {{Solvers}} for geophysical fluid dynamics problems in periodic domains on {{CPUs}} \& {{GPUs}}}.  \jt{J. Open Source Softw.}  \bvol{6}~(60),  \pg{3053}.

\bibitem[Eaves {\em et~al.\/}(2024)Eaves, Maddison, Marshall \& Waterman]{eavesEnergyEnstrophyConstrained2024}
{\sc \au{Eaves, R.~E.}, \au{Maddison, J.~R.}, \au{Marshall, D.~P.} \& \au{Waterman, S.}} \yr{2024}  \bt{An energy and enstrophy constrained parameterization of barotropic eddy potential vorticity fluxes}. Preprint.  \org{Preprints}.

\bibitem[Gallet(2024)]{galletTwodimensionalTurbulenceTopography2024}
{\sc \au{Gallet, B.}} \yr{2024}  \at{Two-dimensional turbulence above topography: Condensation transition and selection of minimum enstrophy solutions}.  \jt{J. Fluid Mech.}  \bvol{988},  \pg{A13}.

\bibitem[Holloway(1992)]{hollowayRepresentingTopographicStress1992}
{\sc \au{Holloway, G.}} \yr{1992}  \at{Representing {{Topographic Stress}} for {{Large-Scale Ocean Models}}}.  \jt{J. Phys. Oceanogr.}  \bvol{22}~(9),  \pg{1033--1046}.

\bibitem[K{\"o}hl(2007)]{kohlGenerationStabilityQuasiPermanent2007}
{\sc \au{K{\"o}hl, A.}} \yr{2007}  \at{Generation and {{Stability}} of a {{Quasi-Permanent Vortex}} in the {{Lofoten Basin}}}.  \jt{J. Phys. Oceanogr.}  \bvol{37}~(11),  \pg{2637--2651}.

\bibitem[LaCasce {\em et~al.\/}(2024)LaCasce, Pal{\'o}czy \& Trodahl]{lacasceVorticesBathymetry2024}
{\sc \au{LaCasce, J.~H.}, \au{Pal{\'o}czy, A.} \& \au{Trodahl, M.}} \yr{2024}  \at{Vortices over bathymetry}.  \jt{J. Fluid Mech.}  \bvol{979},  \pg{A32}.

\bibitem[Mcwilliams(1984)]{mcwilliamsEmergenceIsolatedCoherent1984}
{\sc \au{Mcwilliams, J.~C.}} \yr{1984}  \at{The emergence of isolated coherent vortices in turbulent flow}.  \jt{J. Fluid Mech.}  \bvol{146},  \pg{21--43}.

\bibitem[Mcwilliams(1990)]{mcwilliamsVorticesTwodimensionalTurbulence1990}
{\sc \au{Mcwilliams, J.~C.}} \yr{1990}  \at{The vortices of two-dimensional turbulence}.  \jt{J. Fluid Mech.}  \bvol{219}~(-1),  \pg{361}.

\bibitem[Radko(2023)]{radkoGeneralizedTheoryFlow2023}
{\sc \au{Radko, T.}} \yr{2023}  \at{A generalized theory of flow forcing by rough topography}.  \jt{J. Fluid Mech.}  \bvol{961},  \pg{A24}.

\bibitem[Rhines \& Young(1982)]{rhinesHomogenizationPotentialVorticity1982}
{\sc \au{Rhines, P.~B.} \& \au{Young, W.~R.}} \yr{1982}  \at{Homogenization of potential vorticity in planetary gyres}.  \jt{J. Fluid Mech.}  \bvol{122}~(-1),  \pg{347}.

\bibitem[Salmon {\em et~al.\/}(1976)Salmon, Holloway \& Hendershott]{salmonEquilibriumStatisticalMechanics1976}
{\sc \au{Salmon, R.}, \au{Holloway, G.} \& \au{Hendershott, M.~C.}} \yr{1976}  \at{The equilibrium statistical mechanics of simple quasi-geostrophic models}.  \jt{J. Fluid Mech.}  \bvol{75}~(4),  \pg{691--703}.

\bibitem[Santangelo {\em et~al.\/}(1989)Santangelo, Benzi \& Legras]{santangeloGenerationVorticesHighresolution1989}
{\sc \au{Santangelo, P.}, \au{Benzi, R.} \& \au{Legras, B.}} \yr{1989}  \at{The generation of vortices in high-resolution, two-dimensional decaying turbulence and the influence of initial conditions on the breaking of self-similarity}.  \jt{Phys. Fluids A}  \bvol{1}~(6),  \pg{1027--1034}.

\bibitem[Siegelman \& Young(2023)]{siegelmanTwodimensionalTurbulenceTopography2023}
{\sc \au{Siegelman, L.} \& \au{Young, W.~R.}} \yr{2023}  \at{Two-dimensional turbulence above topography: {{Vortices}} and potential vorticity homogenization}.  \jt{PNAS}  \bvol{120}~(44),  \pg{e2308018120}.

\bibitem[Solodoch {\em et~al.\/}(2021)Solodoch, Stewart \& McWilliams]{solodochFormationAnticyclonesTopographic2021}
{\sc \au{Solodoch, A.}, \au{Stewart, A.~L.} \& \au{McWilliams, J.~C.}} \yr{2021}  \at{Formation of {{Anticyclones}} above {{Topographic Depressions}}}.  \jt{J. Phys. Oceanogr.}  \bvol{51}~(1),  \pg{207--228}.

\bibitem[Wang \& Stewart(2018)]{WangStewart2018Slope}
{\sc \au{Wang, Y.} \& \au{Stewart, A.~L.}} \yr{2018}  \at{Eddy dynamics over continental slopes under retrograde winds: Insights from a model inter-comparison}.  \jt{Ocean Model.}  \bvol{121},  \pg{1--18}.

\bibitem[Zhang \& Xie(2023)]{zhangSpectralCondensationQuasigeostrophic2023}
{\sc \au{Zhang, L.-F.} \& \au{Xie, J.-H.}} \yr{2023} Spectral condensation in quasi-geostrophic turbulence above small-scale topography,  \arxiv{arXiv: 2311.16612}.

\end{thebibliography}

\end{document}